# Upper critical field, lower critical field and critical current density of FeTe$_{0.60}$Se$_{0.40}$ single crystal

C S Yadav and P L Paulose

Department of Condensed Matter Physics and Material Sciences, Tata Institute of Fundamental Research, Colaba, Mumbai - 400005 (India).
E mail: csyadav@tifr.res.in, Paulose@tifr.res.in

**Abstract.** The transport and magnetic studies are performed on high quality FeTe$_{0.60}$Se$_{0.40}$ single crystals to determine the upper critical field ($H_{c2}$), lower critical field ($H_{c1}$) and the critical current density ($J_c$). The value of the upper critical field $H_{c2}$ is very large, whereas the activation energy determined from the slope of the Arrhenius plots is found to be lower than that in the FeAs122 superconductors. The lower critical field is determined in 'ab' direction and 'c' direction of the crystal, and found to have anisotropy $\Gamma$ (= $H_{c1//c}$ / $H_{c1//ab}$) ~ 4. The magnetic isotherms measured up to 12 Tesla shows the presence of fishtail behavior. The critical current density at 1.8K of the single crystal is found to be almost same in both 'ab' and 'c' directions in the low field regime.

## 1. Introduction

The discovery of superconductivity in the Fe based oxypnictide compounds has enriched and opened up newer horizons in the field of superconductivity [1]. The tetragonal compounds FeSe and FeTe$_{1-x}$Se$_x$ have relatively simpler structure than the FeAs based superconductors, where the Fe(Te/Se) layers stack along the c axis, and has transition temperature ($T_c$) as high as 15K [2,3,4,5,6,7,8,9]. Pressure studies on the FeSe compounds show an increase in the $T_c$ up to 36K at 38GPa pressure [2,10,11]. Though the $T_c$ in these compounds is much less compared to the FeAs based superconductors, the simplicity of structure and similarity in the Fermi surface make them a potential material to understand the superconducting mechanism in the Fe based oxypnictides. The Fermi surface of the FeS, FeSe and FeTe is very similar to that of FeAs based superconductors, with the cylindrical hole and electron sections at the center and the corner of the brillouin zone respectively [12]. The end member FeTe in the FeTe$_{1-x}$Se$_x$ series is antiferromagnetic below 65K and shows a simultaneous structural transition [4,5,8,9]. Among the Fe mono-chalcogenide compounds, only FeSe shows superconductivity ($T_c$ = 8K), but it is difficult to prepare FeSe in pure form as 1-2% impurity of Fe$_7$Se$_8$ hexagonal phase forms along with the superconducting tetragonal FeSe phase [2,4,6,7,8]. The substitution of Te at the Se site in FeSe increases the $T_c$, showing a maximum close to 40% Se concentration [4,5,9].

In this paper we have estimated the upper critical field ($H_{c2}$), activation energy ($U_0$), lower critical field ($H_{c1}$), and the critical current density ($J_c$) of a high quality single crystal of FeTe$_{0.60}$Se$_{0.40}$ with more than

95% superconducting volume fraction. We have also observed the fishtail behavior in the high field magnetization loop at temperatures below $T_c$ for both directions.

## 2. Experimental Methods

The single crystal of $FeTe_{0.60}Se_{0.40}$ compound were prepared by the chemical reaction of the elements (Fe chunk of 99.999% purity, Te powder of 99.99% purity and Se powder of 99.98% purity) in the stochiometric proportion, inside a sealed quartz tube under vacuum. The charge was slowly heated to $950^0C$ at the rate $50^0C$/hrs and kept for 12 hours before cooling down to $400^0C$ at the rate of $6^0C$/hrs, and then furnace cooled to room temperature for growing the crystals.

The magneto-transport measurements were done using a Quantum Design PPMS (Physical Properties Measurement System). The Specific heat of crystal was measured using relaxation technique in the PPMS. AC susceptibility measurements, and low field DC magnetization was carried out using a SQUID magnetometer and high magnetic field measurements were done using a 12 Tesla Vibrating Sample Magnetometer (Oxford Instruments).

## 3. Results and Discussions

The crystals were found to be very shiny, grown along the 'ab' plane, and were very easy to cleave along this plane. The X-Ray Diffraction (XRD) analysis done on the powdered sample, confirmed the compound to be in their single tetragonal phase (space group P4/nmm), with the lattice parameters *'a' = 3.798Å* and *'c'= 6.058Å*. The compositional analysis by EDAX (Energy Dispersive Absorption X ray Spectroscopy), showed the crystals to be formed in the stochiometric ratio. XRD pattern of the crystal flake shows peaks only at the angles corresponding to the {00l} planes, confirming the orientation of the flakes along the ab plane. To further check the quality of crystal, the TEM (Transmission Electron Microscope) diffraction pattern (shown in the figure 1) was taken, which again confirmed the tetragonal phase and, growth of crystal along the ab-plane.

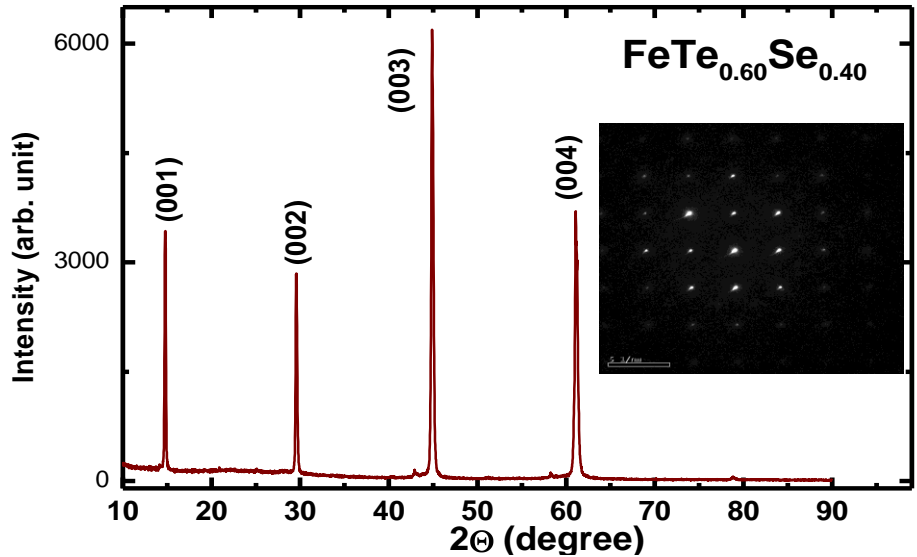

**Figure 1.** XRD diffraction pattern shows only *00l* plane, indicating the growth of crystal along ab-plane. Inset of the figure shows the electron diffraction pattern of $FeTe_{0.60}Se_{0.40}$.

In the Figure 2, we have shown the resistivity data of $FeTe_{0.60}Se_{0.40}$ single crystals for magnetic field parallel to 'c' and the electrical current in the 'ab' plane. The room temperature resistivity is about 0.9mΩ-cm. It is metallic below about 150K and superconducts with a $T_c$ onset of 15.3K (inset of figure 2). At zero magnetic field, the transition width is 0.5K, which is considerably broadened to 2.3K at 14T field. However the $T_c$ onset is not affected very much by the magnetic field as reported in the case of cuprate superconductors. Like the two dimensional cuprate superconductors, the FeAs based layered systems are also reported to have very high critical field [13]. In the figure 3, we have plotted the H-T phase diagram for the crystal corresponding to the temperatures where the resistivity drops to the 90% of the normal state resistivity $\rho_n$, (where $\rho_n$ is taken at temperature T = 16K), 50% of $\rho_n$ and 10 % of $\rho_n$. Since the transition temperature does not shift much towards the low temperatures, it indicates to a very high value of $H_{c2}(0)$ at zero temperature. The linear extrapolation of the lines on field axis at T = 0K, gave the values of high critical field $H_{c2}(0)$ as 184T, 88T and 69T corresponding to the transition temperature taken at the point of 90% of $\rho_n$, 50% of $\rho_n$ and 10 % of $\rho_n$ respectively. Using the Werthamer – Helfand - Hohenberg (WHH) formula

$$\mu_0 H_{c2}(0) = -0.693 \mu_0 \left(\frac{d H_{c2}}{d T}\right)_{T_c} T_c$$

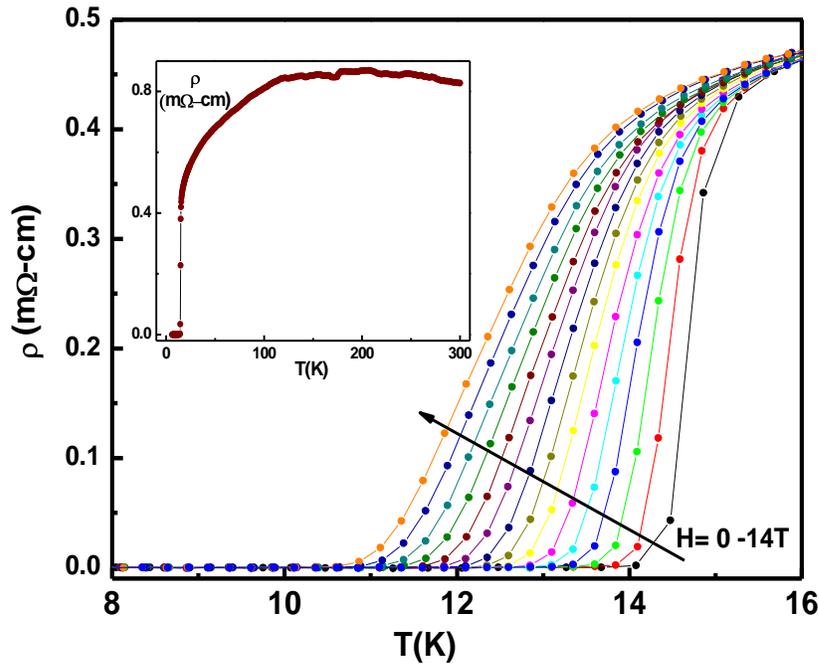

**Figure 2.** Temperature dependence of the resistivity of $FeTe_{0.60}Se_{0.40}$ single crystal, measured in the magnetic fields (from right to left) H = 0, 1, 2, 3, 4, 5, 6, 7, 8, 9, 10, 11, 12, 13, and 14Tesla. The semi-metallic behavior of resistivity in the normal region is shown in the inset of figure.

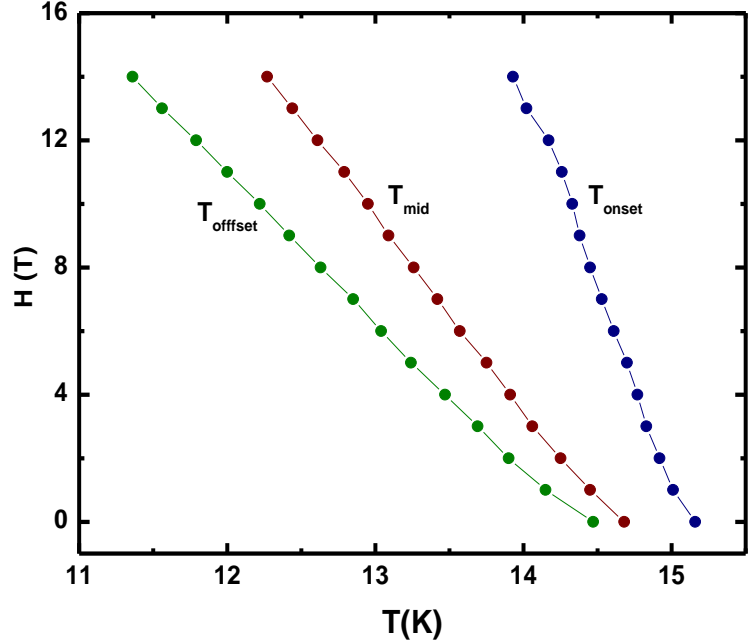

**Figure 3.** Upper critical field versus temperature phase diagram is shown for the points where electrical resistivity drops to 90% of $\rho_n$, 50% of $\rho_n$ and 10% of $\rho_n$, shown by $T_{onset}$ and $T_{mid}$ and $T_{offset}$. $\rho_n$ is the value of resistivity taken in the normal state at 16K.

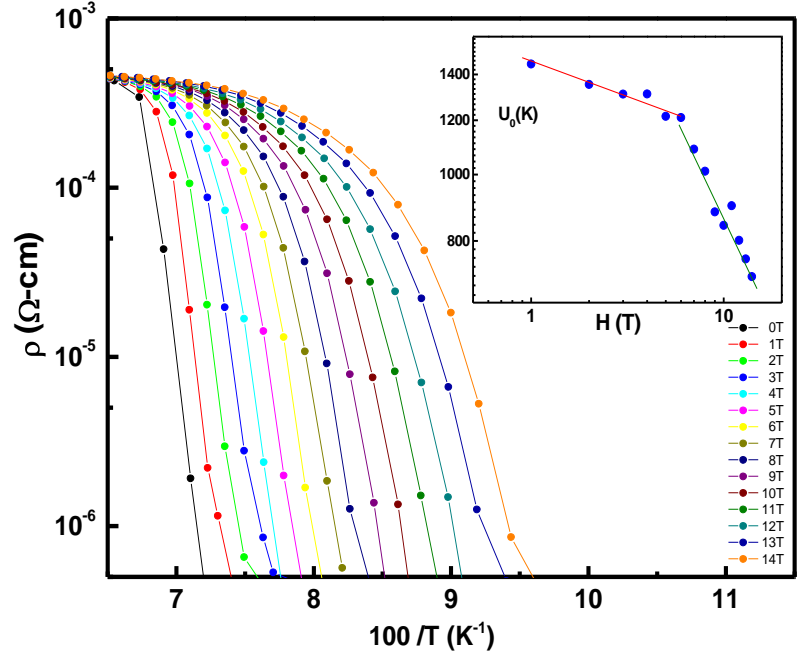

**Figure 4.** Arrhenius plot of the resistivity for FeTe$_{0.60}$Se$_{0.40}$ single crystal for (left to right) H = 0, 1, 2, 3, 4, 5, 6, 8, 9, 10, 11, 12, 13, and 14Tesla. The inset of the figure shows variation of $U_0$ with magnetic field.

to the H-T phase diagram shown in figure 3, the H$_{c2}$(0) were found to be 126T, 65T and 51T corresponding to the points 90% of $\rho_n$, 50% of $\rho_n$ and 10 % of $\rho_n$ respectively. Using these zero temperature value of H$_{c2}$, the corresponding value of $\mu_0 H_{c2}/k_B T_c$ comes out to be 8.21$T/K$, 4.26$T/K$ and

3.30$T/K$, which are much higher than the Pauli limit for $\mu_0 H_{c2}/k_B T_c = 1.84 T/K$ in case of singlet pairing and weak spin orbit coupling [13,14]. This indicates toward the unconventional nature of the superconductivity. In order to get a rough idea about the superconducting parameters, we have used the Ginzburg-Landau (GL) formula for the coherence length ($\xi$), $\xi = (\Phi_0 / 2\pi\mu_0 H_{c2})^{1/2}$, where $\Phi_0 = 2.07 \times 10^{-7}$ Oe cm$^2$, the coherence length $\xi$ at the zero temperature was calculated as 16.2Å, 22Å and 25.5Å for the $H_{c2}$ at 90% of $\rho_n$, 50% of $\rho_n$ and 10% of $\rho_n$ respectively.

The Arrhenius plot for the FeTe$_{0.60}$Se$_{0.40}$ in the figure 4, shows that the electrical resistivity is thermally activated in the region of resistivity between $2 \times 10^{-4}$ $\Omega$-cm and $2 \times 10^{-6}$ $\Omega$-cm. The activation energy $U_0$ is determined from the slope of the curve in this linear region using the formula $\rho(T, H) = \rho_0 \exp(-U_0/k_B T)$. The magnetic field versus the activation energy $U_0$ plot shown in the inset of the figure 4 suggests the different power law dependence on magnetic field $U_0 \propto H^\alpha$, with $\alpha = 0.10$ for $0 < H < 6T$ and $\alpha = 0.57$ for $6T < H < 14T$. Similar kind of power law dependence has also been observed for other superconducting compounds viz. Bi$_2$Sr$_2$CaCu$_2$O$_{8+\delta}$, MgB$_2$, SmFeAsO$_{0.9}$F$_{0.1}$, and NdFeAsO$_{0.82}$F$_{0.18}$ [15,16,17,18]. The activation energy varies linearly from 710K to 1490K for the magnetic field of H = 14T and H = 0T, respectively.

The temperature dependence of the specific heat and the AC magnetic susceptibility of FeTe$_{0.60}$Se$_{0.40}$ are shown in the figure 5(a). The inset of this figure contains the M-H hysteresis loop measured at 1.8K temperature. Though the AC susceptibility data shows almost complete expulsion of the magnetic flux in the Meissner state (~ 95% of the superconducting volume), the kink in the specific heat is not very sharp as expected in the case of second order superconducting transition [19]. The Meissner value of the diamagnetic susceptibility is almost constant below 10K. The value of AC susceptibility in the region 1.8-10K, is almost constant at $\chi = -1.35 \times 10^{-2}$ emu/gm.

We measured the field dependence of the magnetization in the superconducting state at different temperatures, with the external magnetic field along the ab-basal plane and c-axis. The figure 5(b) shows the different magnetization isotherms for the field direction along the ab- plane. The linear variation of the magnetization (-M), which is the signature of the Meissner state is clearly seen in the low field region. The lower critical field $H_{c1}$, as determined from the point of deviation from the linear M versus H of the DC magnetization data was calculated to be 82 Oe for H < ab at T = 1.8K. The low field slope of the M-H curve shown as a dotted line, gives the value of susceptibility $\chi_{dc} = 1.15 \times 10^{-2}$ emu/gm, which is very near to the value of AC susceptibility $\chi_{ac} = 1.35 \times 10^{-2}$ emu/gm at 1.8K (also shown in figure 5(b) in the Meissner state.

For the accurate determination of lower critical field we substracted the value of magnetization (M$_0$) obtained by the low field magnetization slope, from the magnetization (M) for each isotherm. [20,21] The deviation point of $\Delta M$ (i.e. M - M$_0$) versus field curve from the zero base line gives the value of first penetration field ($H_{c1}^*$), where the vortex starts entering in to the sample. The $\Delta M$ versus H plots thus obtained for diffferent temepratures, are shown in the figure 5(c). The inset of the plot show the show $\Delta M$ versus H for T =1.8K. The lower critical field $H_{c1}$ can be deduced from $H_{c1}^*$. For a rectangle sample geometry, Brandt gave the relation between $H_{c1}$ and $H_{c1}^*$ as $H_{c1}^* = H_{c1}/\tanh\sqrt{(0.36\, b/a)}$, where 'a' and 'b' are the width and the thickness of the samples, respectively [22]. Using this formula we estimated effective demagnetizing factor $N_{eff} \sim 0.79$ for our sample of dimension 'a' = 2.2mm and 'b' = 0.25mm.

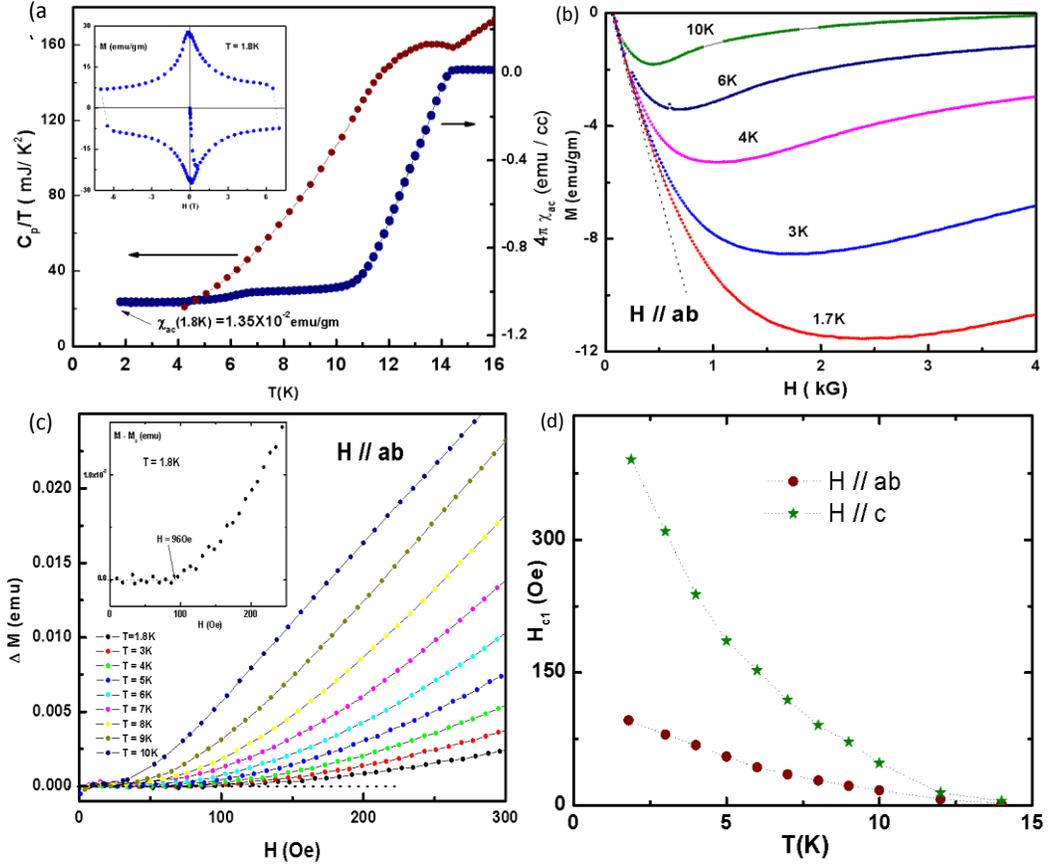

**Figure 5.** (a) The temperature dependence of the specific heat and the ac susceptibility of the FeTe$_{0.60}$Se$_{0.40}$ crystal. The inset shows M-H loop at 1.8K. (b) Field dependence of the initial magnetization isotherms is plotted for different temperatures. Dotted line gives the linear fit to the low field MH curve at 1.8K. (c) Deviation of M, from the linear low field M-H slope ($\Delta$M) is plotted for different temperatures. Inset of figure (c) shows the point of deviation of $\Delta$M from zero base line for T =1.8K, giving the value of first penetration field.(d) Lower critical field H$_{c1}$ measured for H // c and H // ab shows a positive curvature all along temperature region. The Anisotrpy parameter $\Gamma$ ( = $(H_{c1}//c)/(H_{c1}//ab)$) were estimated to be ~ 4 at 1.8K.

As shown in the figure 5(d), the H$_{c1}$ values for H //c and H //ab are highly temperature dependent and show upward trend with negative curvature. Similar trend is reported for FeAs based superconductors viz. Ba$_{0.60}$K$_{0.40}$Fe$_2$As$_2$ and SmFeAsO$_{0.9}$F$_{0.1}$ also. This has been pointed as not conforming with the single band gap description of the mean field theory, and hence as evidences of two energygaps like the MgB$_2$ superconductor [23,24,25]. The density functional study of the FeSe and FeTe done by Subedi et.al. showed that the band structure of these copounds consists cyndirical electron fermi surface at the zone corner and two concentric cynderical hole surface a the zone center, indicating that the superconductivity in this system results from two bands [12], and the upward curvature of H$_{c1}$ is dictated by both electrons and the heavy holes. The H$_{c1}$ values for our FeTe$_{0.60}$Se$_{0.40}$ crystal were found to have an anisotpy ratio ($\Gamma$ = $(H_{c1}//c)/(H_{c1}//ab)$) of 2 - 4 for temperature range 1.8K < T < 14K. This anisotropy is large compared to that in PrFeAsO$_{1-y}$ and Sm$_{0.95}$La$_{0.05}$FeAsO$_{0.85}$F$_{0.15}$ [26,27].

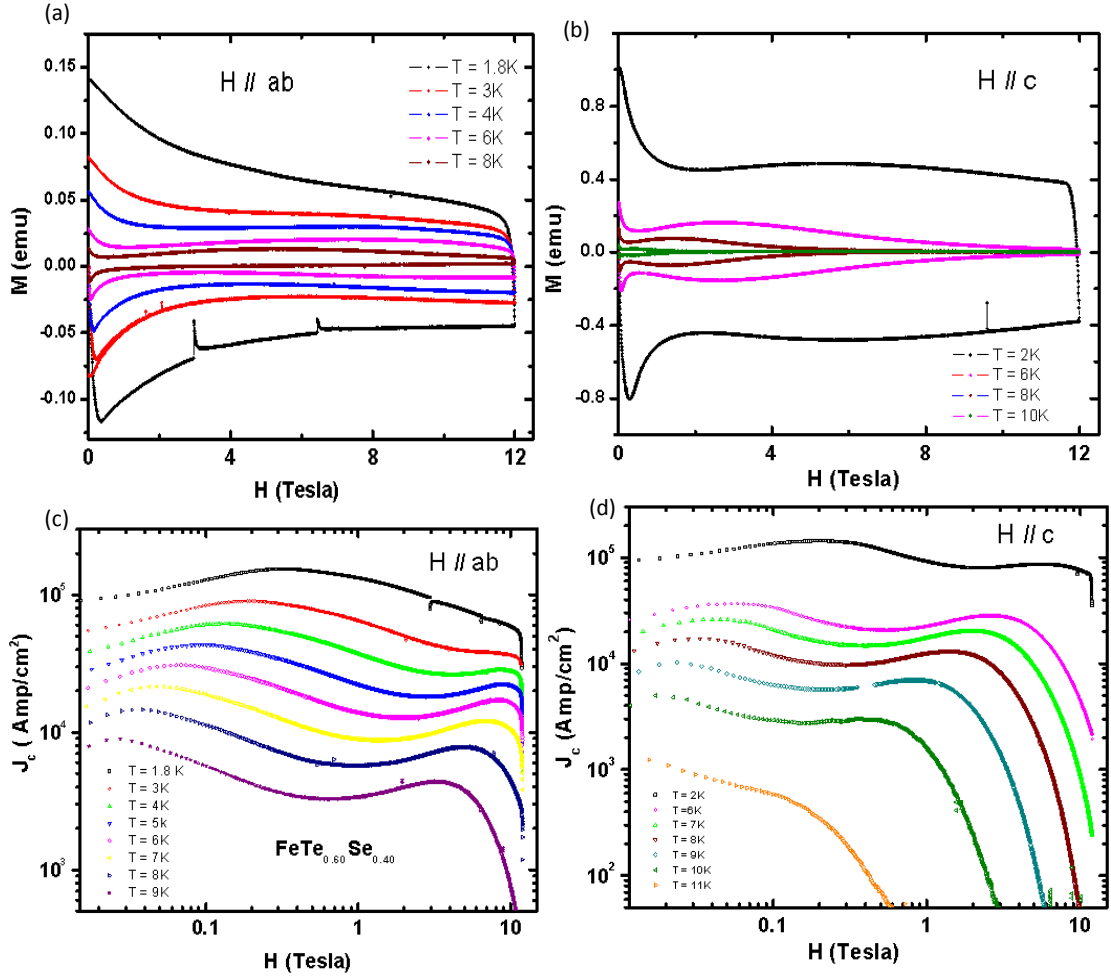

**Figure 6.** Field dependence of magnetic isotherms measured up to 12 Tesla field for (a) H //ab plane and (b) H //c axis, shows the fishtail like feature at higher temperatures. The field dependence of the Critical current density $J_c$ at different temperatures is plotted on log-log scale for (c) H // ab plane and (d) H //c axis.

Figure 6(a, b) shows the M-H loop in positive field direction at several temperatures in the magnetic field parallel to ab-plane (H//ab) and parallel to c axis (H//c), which was measured up to 12 Tesla. The magnetization '−M' goes through a first maximum on increasing the magnetic field and shows a second peak before it finally collapses to zero near the upper critical field $H_{c2}$. This second maximum is known as fishtail effect in the literature and has also been observed for crystals of LaSrCuO, YBCO, BSCCO and more recenlty in the $Ba(Fe_{0.93}Co_{0.07})_2As_2$ single crystals [28,29,30,31]. Though the origin of this behavior is not fully explained yet, one model correlates it to the presence of some weakly superconducting or non superconducting regions which can act as the efficient pinning centers [29,30]. It is also propounded that the crossover from the single to collective flux creep induces a slower magnetic relaxation at the intermediate field and give rise to the second peak [29,30,31]. However the fishtail is strongly dependent on the sample orientation of the externally applied field, and for H parallel to the ab plane this feature get diminished.

Using the Bean's model for the field independent critical current density ($J_c$), it can be calculated by the relation [32,33]

$$J_c = \frac{20.\Delta M}{a\left(1 - \frac{a}{3b}\right)}$$

where $\Delta M$ is $M_{up} - M_{dn}$ and $M_{up}$ and $M_{dn}$ are the magentization while decreasing and increasing magnetic field respectively; a, b are the sample width (a < b). We took sample dimensions as a = 2.2mm , b = 3mm and a = 0.25mm, b =2.2mm for the $J_c$ calcualtion for H // ab and H // c respectively. It should be mentioned that though the Bean's model is strictly applicable only in the case of field independent critical current density, since the variation of $J_c$ is moderate upto 6T for H//ab (upto 1T for H//c), it serves as a good approximation to the actual value.

The critical current density $J_c$ obtained for the FeTe$_{0.60}$Se$_{0.40}$ single crystal sample for H // ab and H // c are shown in the figure 6(c, d). The value of $J_c$ at low field and 1.8K temperature are almost same as $1 \times 10^5$ Amp/cm$^2$, for both directions. The fishtail feature is also more clearly evident. Our value of $J_c$ agrees with the one recent report by Taen et.al. for the crystal of FeTe$_{0.61}$Se$_{0.39}$. [33] However in earlier report for Fe$_{1+y}$Te$_{1-x}$Se$_x$; x = 0.133, Rongwei Hu et.al. have reported an anisotropy in critical current density ($J_{c\;//ab}/J_{c\;//c}$) ~5, in their single crystal with 10% superconducting volume fraction [34]. The current density $J_c$ values also compare well with for the Co doped BaFe$_2$As$_2$ supercondcutor [35].

## 4. Conlcusions

We have determined the upper critical field (H$_{c2}$), activation energy ($U_0$), lower critical fields (H$_{c1}$) and the critical current density ($J_c$) of the FeTe$_{0.60}$Se$_{0.40}$ single crystal. The H$_{c2}$ value at T = 0K measured along the ab plane, from the extrapolation of H-T phase diagram and also using WHH formula are found to be very high. The activation energy shows linear dependence with the magnetic field. The H-T phase diagram for H$_{c1}$ shows a positive curvature and does not saturate till 1.8K. The lower critical field was found to be anisotropic with the anisotropy parameter Γ ( = (H$_{c1\;//c}$)/(H$_{c1\;//ab}$)) ~ 4 at 1.8K. The high field M-H behavior shows the fishtail behavior and is more pronounced for H // c direction. The critical current density $J_c$ of the compound is found to be $1 \times 10^5$ Amp/cm$^2$ at low field and 1.8K temperature, and appears to be isotropic in nature.

**Acknowledgement**

We would like to acknowledge Manish Ghagh for his help with sample preparation and measurements.